\documentstyle[preprint,floats,tighten,prd,aps]{revtex}
\input epsf.tex
\def\DESepsf(#1 width #2){\epsfxsize=#2 \epsfbox{#1}}
\begin{document}
\preprint{\vbox{
\hbox{OITS 607}
\hbox{SLAC-PUB 7218-T/E} 
}}
\draft

\title{Probing the QCD pomeron in $e^+e^-$ collisions}
\author{S.J.\ Brodsky$^{a}$, F.\ Hautmann$^{b}$ and D.E.\ Soper$^{b}$}
\address{$^{a}$ Stanford Linear Accelerator Center, 
Stanford University, Stanford, CA 94309}
\address{$^{b}$ Institute of Theoretical Science, 
University of Oregon, Eugene, OR 97403}
\date{10 December 1996; revised version 14 August 1997}
\maketitle

\begin{abstract}
The total cross section for scattering virtual photons at high energy
is sensitive to pomeron physics.  If the photons are sufficiently
virtual, QCD perturbation theory applies, so that the reaction probes
the short distance pomeron.  We study this reaction for present and
future $e^{\pm} e^{-}$ colliders.
\end{abstract}


%****************************************************

%\narrowtext

%23456789012345678901234567890123456789012345678901234567890123456789012

The  behavior of scattering in the  limit of high energy and fixed
momentum transfer is one of the outstanding open questions  in the
theory of strong interactions.  A useful description of this
behavior was developed in the 1960's, based on analytic continuation
in the complex $J$ plane, where $J$ represents the angular momentum
of exchanged particles \cite{Regge}. In the late 1970's, Lipatov and
collaborators established a connection of the high energy behavior
with the quark and gluon degrees of freedom, at least for special
situations in which perturbation theory applies \cite{FKL}. These 
papers still form the core of our knowledge of QCD scattering  at
high energies.   The physical effect they describe is often referred
to  as  the  QCD pomeron, or BFKL effect.

Experimental studies of the QCD pomeron are at present carried out
mainly   at the HERA $e \, p$ collider  in deeply inelastic scattering
in the region of low values of the Bjorken variable $x$
\cite{herarev}. In this case, QCD pomeron effects are expected to
give rise to a power-law growth of the structure functions  as $x$
goes to zero, and an increase in the scaling violation  at fixed
values of $x$ (for small enough $x$).  However, studying BFKL pomeron
effects in deeply inelastic scattering   is made difficult  by the
fact that the low $x$ behavior is influenced by both short  distance
and long distance physics. As a result,  predictions at photon
virtuality $Q$ depend on a set of non-perturbative inputs  at a scale
$Q_0<Q$.  This makes it difficult  to untangle perturbative BFKL
predictions from non-perturbative effects.

Other processes have been suggested in which the ambiguity related to
large distance physics is expected to be reduced. One such process is
the production of two jets in hadron-hadron collisions
\cite{navelet}. One looks for jets produced by inelastic
parton-scattering at fixed momentum $p_T$  transferred  to the
parton  in the limit of large parton energy $\sqrt{\hat s}$. In order
to  avoid entanglement with the parton distributions, one   has to 
keep $\hat s/s$ fixed as $\hat  s \to \infty$.  Unfortunately, this is
not readily done at a single hadron collider. An essentially
equivalent process is to look for a high $p_T$ jet with a large
fraction of the proton's longitudinal momentum in the final state of a
deeply inelastic scattering event at small $x$.

A related possibility is to look for elastic parton-scattering at
large $\hat s$ with fixed $p_T$ with the  demand that the exchanged
quanta be in a color singlet state. The signature for this is the
existence of a gap in the rapidity interval between the two jets
produced by the scattered partons \cite{bjozep}.  Unfortunately the
gap signal is affected by various long distance processes, including
spectator collisions.

In principle, one could avoid the difficulties mentioned above by
measuring the  cross section for the scattering at large $s$ and fixed
$t$ of two colorless bound states of a heavy quark and a heavy
antiquark (``quarkonia'')  \cite{oniumwave}. This process is
perturbative because of the  small size of the quarkonium state.
Unfortunately, until one can build a quarkonium accelerator, this 
must remain a gedanken experiment.

In this paper  we consider the  possibility of studying BFKL effects
by measuring the total cross section for off-shell photons at 
$e^{\pm} \, e^{-}$ colliders.  Essentially, we replace the quarkonia
by virtual photons.  The size of the wave function for finding 
quarks in a virtual photon is controlled by the photon virtuality
instead of the heavy quark mass. An exciting feature of such an
experimental investigation would be that the virtualities of each
of the two photons could be tuned by measuring the momenta of the
recoil leptons.

The process that  we discuss, virtual photon scattering, has recently
been used by Balitskii  to reformulate the BFKL problem in terms of an
expansion using Wilson line operators \cite{bali}.  This method may
enable theory  to get beyond the leading logarithm approximation. In
the present paper, we are not concerned  with the foundations of the
theory, but with the practical possibilities  for using virtual photon
scattering as a probe of the short distance pomeron.

There have been other investigations of the high energy regime in the
context  of  $e^{\pm}  e^{-}$  collisions. These have mainly focussed
on photon structure  functions  (see for instance Ref.~\cite{gagaphy}
for a recent overview of this  subject) and  diffractive meson
production \cite{gagaphy,gipom}.   In these cases, either
non-perturbative parton distributions in the photon or
non-perturbative meson wave functions enter the  theoretical
predictions. The results presented in this paper appeared in
preliminary form in \cite{us}. Similar results obtained independently
by Bartels, De Roeck and Lotter may be found in Ref.~\cite{them}. 

We will describe the  total cross section for the scattering of
two  transversely polarized virtual (space-like) photons  $\gamma^{*}
(q_A)$ and  $ \; \gamma^{*} (q_B)$, with virtualities  $q_A^2 = -
Q^2_A$ and $q_B^2 = - Q^2_B$, in the high energy region where $s =
(q_A+q_B)^2$ is much larger than $Q_A^2$ or $Q_B^2$. We suppose
that the photon virtualities are in turn large with respect to the
QCD scale $\Lambda^2_{QCD}$, so that the process occurs at short
distances (much smaller than $\Lambda^{-1}_{QCD} \approx 1 \; {\mbox
{fm}}$) and QCD perturbation theory applies. 

It is simple to see that the two-gluon exchange mechanism gives rise
to a constant $\gamma^*\gamma^*$ total cross section at large $s$, $
\sigma^{(0)}(s,Q_A^2,Q_B^2) \sim \alpha^2 \, \alpha_S^2 \,
f(Q_A^2,Q_B^2)$ \cite{lownus}.  To higher orders in perturbation
theory, the iteration of gluon ladders promotes this constant to 
logarithms, and the perturbative expansion of the cross section at
high energy has the form
\begin{eqnarray}
\label{expansion}
\sigma ({\gamma^{*} \, \gamma^{*}}) &\sim& 
\sigma^{(0)}  \, 
\left[ 1 + \sum_{k=1}^{\infty} a_k \left( \alpha_S \, 
L\right)^k 
+ \dots \right] \,,
\quad\quad\quad
L = \ln (s/Q^2) \,,
\end{eqnarray}
where $Q^2$ is a scale of the order of the initial photon
virtualities, the sum represents the series of the leading logarithms
to all orders  in the strong coupling $\alpha_S$, and the dots stand
for non-leading terms.  To study the high energy behavior, it is  
convenient to  analyze the cross section in its Mellin moments, 
defined by 
\begin{eqnarray}
\sigma (s, Q_A^2, Q_B^2) &=& \int_{a - i \infty}^{a + i \infty} 
\, { { d \, N} \over { 2  \pi  i} } \,  
e^{ N \, L}
\sigma_N (Q_A^2, Q^2_B)\,,
\label{mellin}
\end{eqnarray}
where the $N$-integral runs parallel to the imaginary axis and to the
right of  any singularities in $\sigma_N$. We see from this definition
that a constant behavior of the cross section with the energy $s$ is
generated by a simple pole in $\sigma_N$ at $N = 0$, while powers of
logarithms are generated by multiple poles at $N = 0$.

\begin{figure}[htb]
\centerline{ \DESepsf(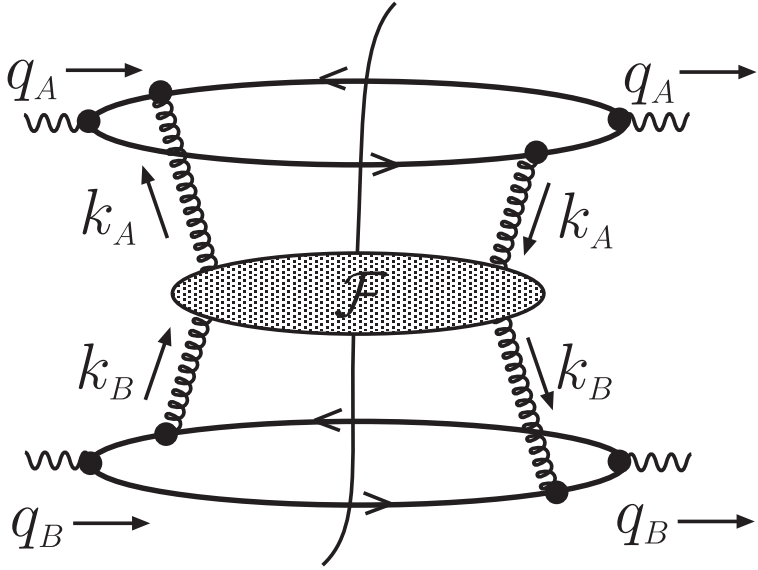 width 10 cm) }
\bigskip
\caption{ Graphical representation of Eq.~(\protect\ref{ktfac}). The
gluons can be attached to the quark lines in 16 different ways.}
\label{fig:ktfac}
\end{figure}

To sum the leading logarithmic terms, we employ the method developed
in Refs.~\cite{HEF} for computing hard scattering cross sections at
high energy. As illustrated in Fig.~\ref{fig:ktfac}, we write
$\sigma_N$  as a convolution of three factors,
\begin{eqnarray}
\sigma_{ N } ( Q_A^2, Q_B^2)  &=&  
\int \, {{d^2 \, {{\bf k}}_A} \over { \pi \, {{\bf k}}^2_A}} \,
\int \, {{d^2 \, {{\bf k}}_B} \over { \pi \, {{\bf k}}^2_B}}
\
G({{\bf k}}^2_A/ Q^2_A) \, {\cal F}_N  ({{\bf k}}_A, {{\bf k}}_B) \, 
G({{\bf k}}^2_B/ Q^2_B) \, . 
\label{ktfac}
\end{eqnarray}
The factors $G$ describe the coupling of two gluons to a
quark-antiquark pair created by the virtual photon.  We compute $G$ at
lowest (one loop) order, taking the limit in which the longitudinal
momentum carried by the gluons is negligible compared to the
longitudinal momentum carried by the quarks. At order $\alpha_S^0$,
the other factor, ${\cal F}_N$,  describes the exchange of two gluons
between the $q\bar q$ states:
\begin{equation}
\label{bornfn}
{\cal F}_N^{(0)}  ({{\bf k}}_A, {{\bf k}}_B) = 
{\pi \over N}\, \delta({{\bf k}}_A - {{\bf k}}_B)
\;\;\;\;.  
\end{equation}
The simple pole at $N = 0$ corresponds to a cross section that
is independent of $s$. The full function ${\cal F}_N$ is the
standard solution of the BFKL equation \cite{FKL} that reduces to
(\ref{bornfn}) at order zero.

Inserting the BFKL solution for ${\cal F}_N$ in Eq.~(\ref{ktfac})
and working out the transverse momentum integrations gives
\begin{eqnarray}
{\sigma}_N  ({Q}_A^2, {Q}_B^2)
&=& 
{ 1 \over 2 \, \pi \, {Q}_A \, {Q}_B  } \, 
\int_{1 / 2 - i \infty}^{1 / 2 + i \infty} \, 
{{d \, \gamma} \over {2  \pi  i}}
\
 \left( {{  {Q}^2_A } \over { {Q}^2_B} } \right)^{\gamma - 1
/ 2 } 
\, {V(\gamma) \, V ( 1 - \gamma) \over  {N - {\bar \alpha_S} \, \chi
(\gamma) }}  
\, , 
\label{gareprsi}
\end{eqnarray}
where we have set $ {\bar \alpha_S} =   \alpha_S \, C_A 
/ \pi$ ($C_A = 3$), and the functions $\chi$ and $V$ are given by 
\begin{equation} 
\label{chi}
\chi (\gamma) = 2\, \psi (1 ) - \psi( \gamma) - \psi (1 - \gamma) 
\hspace*{0.8 cm} , 
\end{equation}
\begin{eqnarray}
V ( \gamma) &=&  \pi \, \alpha \, \alpha_S \,
 \left( \sum_q e^2_q \right)  \,
\
{ { (1 + \gamma) \, ( 2 - \gamma) \, 
\Gamma^2 (\gamma) \, \Gamma^2 ( 1 - \gamma) } \over 
{(3 - 2 \, \gamma) \, \Gamma ( 3 / 2  + \gamma)
\, \Gamma ( 3 /2  - \gamma) }}\,.
\label{viexp}
\end{eqnarray}
Here $\Gamma$ is the Euler $\Gamma$-function, and $\psi$ is its
logarithmic  derivative. While the function $\chi(\gamma)$ is a
universal function from the solution to the BFKL equation,  the
function $V(\gamma)$ is  specific to the off-shell photon probe. 

\begin{figure}[htb]
\centerline{ \DESepsf(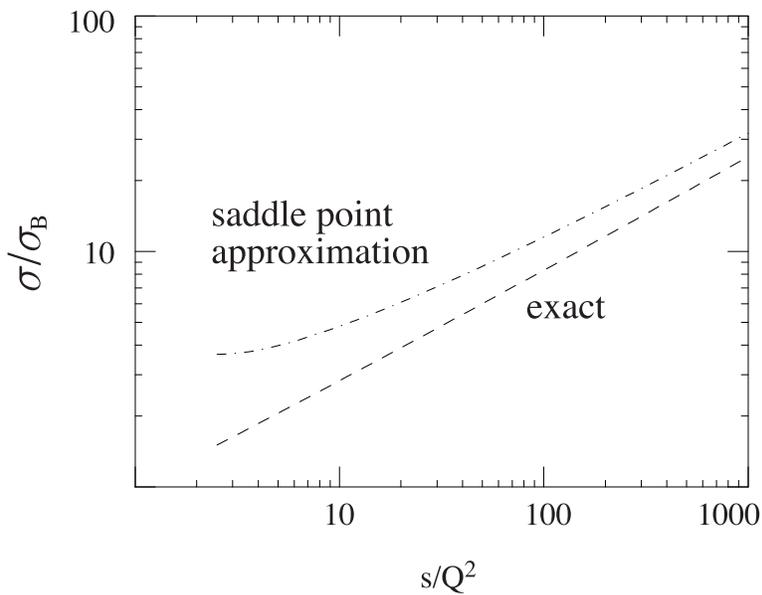 width 12 cm) }
\bigskip
\caption{ The $s/Q^2$ dependence of the  $\gamma^{*} \, \gamma^{*}$
total cross section, Eq.~(\protect\ref{sigofs}). We take  $Q_A^2 =
Q_B^2$,  $\alpha_S = 0.2$, and divide $\sigma$ by the Born cross
section $\sigma_B$, which is independent of $s$.  We show the results
of evaluating the integral exactly and of using the saddle point
approximation.}
\label{fig:xsect}
\end{figure}

Eq.~(\ref{gareprsi}) sums the $1/N$ poles to the  accuracy
$(\alpha_S^2/N)\times\left( \alpha_S / N \right)^k $, for any $k$. By
evaluating Eq.~(\ref{gareprsi}) to the lowest perturbative order, $k
= 0$, one recovers the constant contribution $\sigma^{(0)}$ from the
two-gluon exchange model for the pomeron.  In general, one determines
$1/N^{k+1}$ contributions to the cross section by retaining higher
orders in the $\alpha_S$-expansion of Eq.~(\ref{gareprsi}), and we see
that the general structure of the coefficients of the leading
logarithmic series comes from both $\chi (\gamma)$ and $V (\gamma)$.
The net cross section at the leading logarithmic level is obtained
from Eq.~(\ref{gareprsi}) by taking the inverse Mellin transform
(\ref{mellin}):
\begin{eqnarray}
{\sigma} (s,{Q}_A^2, {Q}_B^2) &=& 
{ 1 \over 2 \, \pi \, {Q}_A \, {Q}_B  } \, 
\int_{1 / 2 - i \infty}^{1 / 2 + i \infty} \, {{d \, \gamma} \over 
{2  \pi  i}} \,  
\left( {{  {Q}^2_A } \over { {Q}^2_B} } \right)^{\gamma - 1 / 2 }
\,
\left(  s \over Q^2 \right)^{\bar\alpha_S\,\chi(\gamma)}  
V(\gamma) \, V ( 1 - \gamma)
\, .
\label{sigofs}
\end{eqnarray}
In the high energy limit $s \gg Q^2$, the integral (\ref{sigofs})
is dominated by the region near the saddle point of $\chi(\gamma)$ at
$\gamma = 1/2$. One obtains the BFKL power dependence $s^\lambda$ with
$\lambda = \bar\alpha_S\,\chi(1/2) = 2.77\,\bar\alpha_S$. In
Fig.~\ref{fig:xsect}, we show the cross section computed from
Eq.~(\ref{sigofs}) with $\alpha_S = 0.2$ as a function of $s/Q^2$.

Our primary concern in this paper is with high energy virtual photon
scattering. However, one can also use Eqs.~(\ref{gareprsi}) and
(\ref{sigofs}) to explore small $x$ deeply inelastic scattering (DIS)
\cite{sas}, in which $Q_B^2  \ll Q_A^2 \ll s$. We will investigate
these issues in a future paper \cite{later}.

The cross section for high energy virtual photon scattering can be
measured in $e^{\pm} \, e^{-}$ collisions in which the outgoing
leptons are tagged. The  part of the $e^{\pm} \, e^{-}$ cross section
that is contributed by transversely polarized photons is obtained by
folding the $ \gamma^{*}\, \gamma^{*} $ cross section with the flux
of photons from each lepton:
\begin{eqnarray}
\sigma 
&=&    \int_{\cal R} {{d \, x_A} \over {x_A}} \, 
{{d \, x_B} \over {x_B}} \, 
{{d \, Q^2_A} \over {Q^2_A}} \, 
{{d \, Q^2_B} \over {Q^2_B}} \,
\left( {\alpha \over {2 \, \pi}} \right)^2
x_A P_{\gamma/e^{+}}( x_A) \, x_B P_{\gamma/e^{-}}( x_B) \,
\sigma_{ {\gamma^{*}}{\gamma^{*}}}(x_A  x_B s_{ee}, Q_A^2, Q_B^2) 
.
\label{epatot}
\end{eqnarray}
Here $x_A$ and $x_B$  denote the fractions of the longitudinal momenta
of the leptons $A$ and $B$ that are carried by the corresponding
bremsstrahlung photons. The integration region $\cal R$ is
determined by the experimental cuts. The flux factor is given by
\begin{equation}
\label{gammaflux}
 P_{\gamma/e}( x) =  {{1 + (1 - x)^2 } \over x} \hspace*{0.8 cm} .  
\end{equation}

In the $\gamma^* \gamma^*$ cross section, Eq.~\ref{sigofs}, one
must make choices of the scale $\mu$ in $\alpha_S$ and the variable
$Q^2$ that provides the scale for the logarithms of $s$. A reliable
determination of these scales requires a next to leading order
calculation, which is beyond the scope of this paper. We simply
choose
\begin{eqnarray}
\mu^2&=&c_\mu\, Q_A Q_B
\nonumber\\
Q^2 &=& c_Q\, Q_A Q_B
\end{eqnarray}
with $c_\mu = e^{-5/3}$ and $c_Q = 100$. Our choice for $\mu^2$ is
based on a calculation of the mean logarithmic virtuality for the
gluon propagators in the two gluon exchange graph, following the
prescription of \cite{maclep}. The choice for $Q^2$ is based on an
estimate of the rapidity range available for exchanged gluons. We will
discuss these estimates in a later work \cite{later}. Given the
uncertainties of the leading logarithmic approximation, one should
regard the numerical results that follow as being accurate only at
the order of magnitude level.

Having made these scale choices, one can check Regge factorization:
does the cross section decompose into a product of a function of
$Q_A$ times a function of $Q_B$ times a function of $s$? Looking at
Eq.~(\ref{sigofs}) with $Q^2 = c_Q Q_A Q_B$, one sees that this
factorization is an approximate property of the formula at large
$s$ when the $\gamma$ integral is dominated by a narrow range of
$\gamma$ near the saddle point of $\chi(\gamma)$.

We now estimate the cross section available to study BFKL
effects at $e^+e^\pm$ colliders. We integrate Eq.~(\ref{epatot})
over a range $\cal R$ determined by 1) $Q_A > Q_{\rm min}$, 
$Q_B > Q_{\rm min}$ with $Q_{\rm min}^2 = 5\ {\rm GeV}^2$, in order
to keep $\alpha_S$ small and 2) $x_A\, x_B s_{ee} >
100\  Q_A Q_B$ in order that the high energy approximation be
valid. We base the numerical value of this limit on a demand that
the two gluon exchange graph be dominant over the (lower order)
quark exchange graph when we chose $\alpha_S = 0.2$. 

Performing the integrations numerically and adding a similar
contribution from longitudinally polarized virtual photons, as
described in Ref.~\cite{them}, we obtain 
\begin{equation}
\label{ratevallep}
\sigma  
\simeq 1 \, {\mbox {pb}} 
\hspace*{2 cm}    
( \sqrt{s} = 200 \,     {\mbox {GeV}} )
\end{equation}
at LEP200 energies, and 
\begin{equation}
\label{ratevalnlc}
\sigma   
\simeq 4 \, {\mbox {pb}} 
\hspace*{2 cm}   
( \sqrt{s} = 500 \,     {\mbox {GeV}} ) 
\end{equation}
at a future next linear collider (NLC).  These cross sections would
give rise to about $500$ events at LEP200 for a value of the luminosity
$L = 500 \, {\mbox {pb}}^{-1}$,  and about $2 \times 10^5$ events at
the NLC for $L = 50 \, {\mbox {fb}}^{-1}$.   For $Q_{\rm min}^2 =
36\ {\rm GeV}^2$, corresponding to a minimum electron scattering angle
of 24 mrad, the number of events at the NLC would be about $10^3$.
While this looks rather marginal at LEP200, it appears that measuring
off-shell photon scattering at the NLC could be a viable way of
studying short distance pomeron effects.

Even with a modest luminosity, one can examine experimentally how the
perturbative pomeron emerges from the soft pomeron as $Q_A$ and $Q_B$
are increased. For small $Q_A$ and $Q_B$ a simple Regge model should
apply. For on-shell photons, Regge factorization gives 
\begin{equation}
\sigma_{ {\gamma} {\gamma} } \approx
{ 
\sigma_{ {\gamma} {p} } \;
\sigma_{ {\gamma} {p} } 
/ 
{\sigma_{ p \, p }} 
}\,.
\hspace*{0.8 cm}
\end{equation}
Assuming the values $\sigma_{ {\gamma} \, {p} } \approx 0.1 \; {\mbox
{mb}}$ and $\sigma_{ p \, p } \approx 40 \; {\mbox {mb}}$, one gets
$
\sigma_{ {\gamma} {\gamma} } \approx
250 \; {\mbox {nb}} 
$.
For virtual photons with small $Q_A$ and $Q_B$, the fall-off of the
cross section can be estimated from vector meson dominance:
\begin{equation}
\label{vmd}
  \sigma_{ {\gamma^{*}} {\gamma^{*}} } \approx
\left( {{M^2_\rho} \over 
{{M^2_\rho} + Q_A^2}}  \right)^2 \;\, 
 \left( {{M^2_\rho} \over 
{{M^2_\rho} + Q_B^2}} \right)^2 \;\,
  \sigma_{ {\gamma} {\gamma} }   \hspace*{0.8 cm} . 
\end{equation}

\begin{figure}[htb]
\centerline{ \DESepsf(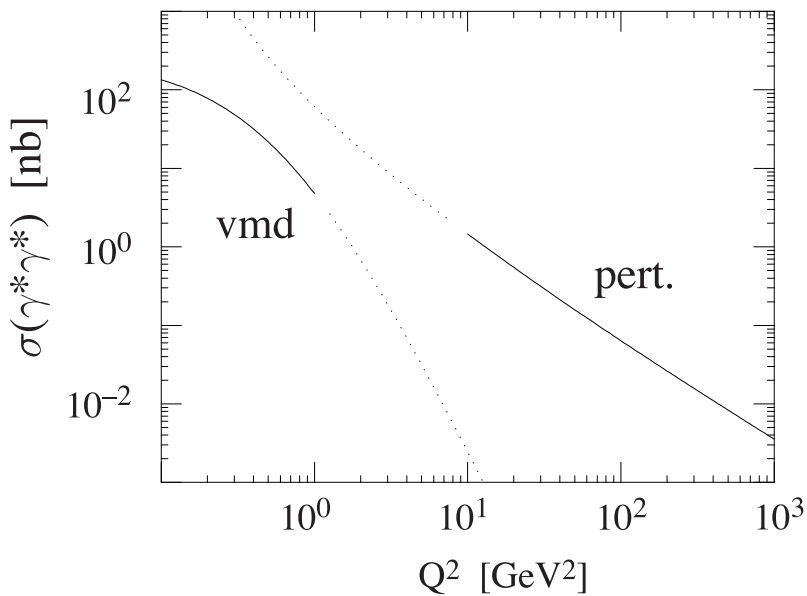 width 12 cm) }
\bigskip
\caption{$Q^2$-behavior of the vector meson dominance and 
       perturbative cross sections in lowest order, with 
       $Q_A^2 = Q_B^2 \equiv Q^2$.}
\label{fig:loglog}
\end{figure}

Fig.~\ref{fig:loglog} shows a log-log plot of the curves corresponding
to the soft and the perturbative formulas (in lowest order) for the
$Q^2$-behavior of the cross section.  In the region $Q^2 \lesssim 1\
{\rm GeV}^2$, one may expect the formula (\ref{vmd}) based on vector
meson dominance to apply. As one goes above this region, the cross
section, instead of continuing to fall like $1/Q^8$,  should begin to
fall more slowly.  At large photon virtualities $Q^2 \gtrsim 10\ {\rm
GeV}^2$ the cross section should exhibit the perturbative scaling
behavior in Eq.~(\ref{gareprsi}), $\sigma \propto 1/Q^2$ at fixed
$(s/Q^2)$.

With a large luminosity, as at a future high energy linear collider,
one can explore virtual photon scattering to higher $Q^2$ and thus
probe experimentally the effects of pomeron exchange in the region
where summed perturbation theory should apply. One should be able to
investigate this region in detail by varying $Q_A$, $Q_B$ and 
$\hat s = x_A x_B s_{ee}$ independently.  

We thank D.\ Strom for useful advice.
This work was supported in part by the United States Department of
Energy grants DE-FG03-96ER40969 and DE-AC03-76SF00515.

\end{document}